\documentclass[]{spie}  

\usepackage{amsmath,amsfonts,amssymb}
\usepackage{graphicx}
\usepackage[colorlinks=true, allcolors=blue]{hyperref}
\usepackage{float}
\usepackage{subcaption}
\usepackage{amsmath}
\usepackage{algorithm}
\usepackage{algpseudocode}
\usepackage[export]{adjustbox}

\DeclareMathOperator*{\argmin}{argmin}
\graphicspath{ {/home/sbhadra/Desktop/SPIE20_ReconGAN/SPIE_20_manuscript/} }

\title{Medical image reconstruction with image-adaptive priors learned by use of generative adversarial networks}

\author[a]{Sayantan Bhadra}
\author[b]{Weimin Zhou}
\author[c]{Mark A. Anastasio}
\affil[a]{Department of Computer Science and Engineering, Washington University in St. Louis, \break St. Louis, MO 63130, USA}
\affil[b]{Department of Electrical and Systems Engineering, Washington University in St. Louis, \break St. Louis, MO 63130, USA}
\affil[c]{Department of Bioengineering, University of Illinois at Urbana-Champaign, \break Urbana, IL 61801, USA}

\authorinfo{Further author information: (Send correspondence to Mark A. Anastasio)\\Mark A. Anastasio: E-mail: maa@illinois.edu, Telephone: 1 217 333 1867}

\begin{document} 
\maketitle

\begin{abstract}
Medical image reconstruction is often an ill-posed inverse problem. In order to address such ill-posed inverse problems, prior knowledge of the sought after object property is usually incorporated by means of regularization.
For example, sparsity-promoting regularization in a suitable transform domain is widely used to reconstruct images with diagnostic quality from noisy and/or incomplete medical data. However, sparsity-promoting regularization may not be able to comprehensively describe the actual prior information of the objects being imaged. Deep generative models, such as generative adversarial networks (GANs) have shown great promise in learning the underlying distribution of images.
Prior distributions for images estimated using GANs have been employed as a means of regularization with impressive results in several linear inverse problems in computer vision that are also relevant to medical imaging. However, in practice, it can be difficult for a GAN to comprehensively describe prior distributions, which can potentially lead to a lack of fidelity between the reconstructed image and the observed data.
Recently, an image-adaptive GAN-based reconstruction method (IAGAN) was proposed to guarantee stronger data consistency by adapting the trained generative model parameters to the observed measurements. In this work, for the first time, we apply the IAGAN method to reconstruct images from undersampled magnetic resonance imaging (MRI) measurements. A state-of-the-art GAN model called Progressive Growing of GANs (ProGAN) was trained on a large number of ground truth images from the NYU fastMRI dataset, and the learned generator was subsequently employed in the IAGAN framework to reconstruct high fidelity images from retrospectively  undersampled experimental k-space data in the validation dataset. It is demonstrated that by use of the GAN-based reconstruction method with noisy and/or incomplete measurements, we can potentially recover fine structures in the object that are relevant for medical diagnosis that may be difficult to achieve using traditional reconstruction methods relying on sparsity-promoting penalties.
\end{abstract}


\section{INTRODUCTION}
\label{sec:intro}  
A linear discrete-to-discrete imaging system is considered \cite{barrett2013foundations}: 
\begin{equation}\label{eq:imaging}
    \textbf{g} = \textbf{Hf} + \textbf{n},
\end{equation}
Here, $\textbf{H}: \mathbb{E}^N \rightarrow \mathbb{E}^M$ denotes the system matrix. The vectors $\textbf{g} \in \mathbb{E}^M$ and $\textbf{n} \in \mathbb{E}^M$ denote the measurement data and random noise, respectively. The vector $\textbf{f} \in \mathbb{E}^N$ represents a finite-dimensional approximation of the measured object's property distribution. Image reconstruction methods seek to estimate the unknown object \textbf{f} from the observed measurement data $\textbf{g}$. Such linear inverse problems are often ill-posed, e.g. when $M << N$ as in the case of compressed sensing MRI applications.

To deal with such ill-posed problems, penalized least squares optimization problems are sometimes solved:
\begin{equation}\label{eq:pls}
    \hat{\textbf{f}} = \argmin_\textbf{f} ||\textbf{g}-\textbf{Hf}||^2_2 + \lambda \Phi(\textbf{f}).
\end{equation}
Here, $\Phi(\textbf{f})$ denotes the regularization or penalty term that encodes the prior on the object, and the hyperparameter $\lambda$ controls the strength of regularization. 
Sparsity-promoting penalties such as the $l_1$-norm of the wavelet transform or the total variation (TV) semi-norm are able to effectively regularize some of these ill-posed linear inverse problems \cite{barrett2013foundations, rudin1992TV, lustig2007CS, tian2011low}.

However, hand-crafted sparsity-promoting penalties may not be able to comprehensively represent the prior knowledge of the sought after object property. 
Recently, deep generative models such as generative adversarial networks (GANs) \cite{goodfellow2014generative} have shown great promise in estimating the prior distributions for images. In the context of medical imaging, such learned priors can be used to perform image reconstruction from incomplete and/or noisy measurement data. Existing approaches use GANs to transform initial images obtained from undersampled measurements using conventional methods, e.g. zero-filled backprojection to artifact-free images while maintaining data consistency \cite{quan2017cyclicGANMRI, mardani2019gancs}. These methods combine learning the prior distribution and the reconstruction tasks together during training. This requires the network to be trained separately each time the data-acquisition parameters are changed.
Bora \textit{et al.} \cite{bora2017csgm} proposed a method in which the training of the GAN and the reconstruction task can be treated separately. In this framework, known as Compressed Sensing using Generative Models (CSGM), a generative model is trained such that the generator can learn to map from simple low-dimensional latent distributions (e.g. uniform, standard normal etc.) to the high-dimensional object distribution. 
The distribution learned by the generator captures the prior knowledge over the object distribution. Subsequently, image reconstruction is performed by finding the latent vector for which the corresponding image in the object space agrees with the observed measurements. Therefore, the GAN needs to be trained only once to learn the prior that describes the object distribution, and the pre-trained generator can be used to reconstruct images from measurements obtained using imaging systems with different data-acquisition parameters.

Still, in practice, it is difficult for a GAN to span all possible images that may arise from the actual distribution. Hence, by constraining the reconstructed image to lie in the range of the generator in the CSGM framework, a potential lack of fidelity may be introduced between the reconstructed image and the observed measurements in the measurement space of the imaging operator $\textbf{H}$ \cite{barrett2013foundations}. 
In order to mitigate the problem of limited representation capabilities of a GAN, Hussein \textit{et al.} \cite{hussein2019IAGAN} proposed an image-adaptive GAN-based (IAGAN) reconstruction framework, where the trained generative model parameters are further tuned to be consistent with the observed measurement data. This results in a higher fidelity with the observed measurements while still maintaining the learned prior over the imaging object obtained by pre-training the GAN. 

In this study, we investigate the application of the IAGAN formulation to image reconstruction in MRI. A state-of-the-art GAN called Progressive Growing of GANs (ProGAN) \cite{karras2018PGGAN} was trained on the publicly available NYU fastMRI dataset (\url{fastmri.med.nyu.edu}) containing knee MRI images and associated k-space measurements. The learned generative model was employed in the IAGAN framework to reconstruct images from highly subsampled k-space data belonging to a previously unseen validation dataset. It is demonstrated that by using an image-adaptive GAN-based reconstruction method with incomplete measurement data, we can obtain high fidelity images and recover fine structures relevant for medical diagnosis, as compared with traditional regularized reconstruction methods that rely on sparsity-promoting penalties.

\section{Image reconstruction with image-adaptive priors learned by use of Generative Adversarial Networks}
\label{sec:sections}
\vspace{0.3cm}
\subsection{Generative Adversarial Networks (GANs)}
Generative Adversarial Networks (GANs) \cite{goodfellow2014generative} are recent deep learning methods that have shown promising results in learning data distributions. In GANs, a generator network and a discriminator network are trained though an adversarial process. 
Here, a true object image $\textbf{f} \in \mathbb{R}^N$ is sampled from a data distribution $p_\textbf{f}$.
The generator maps a random vector $\textbf{z} \in \mathbb{R}^{k}$ to a synthetic object image $\hat{\textbf{f}} = G(\textbf{z}; \theta_{G})$, where $G: \mathbb{R}^{k} \rightarrow \mathbb{R}^N$ is the mapping represented by a neural network with parameters $\theta_G$. The discriminator is an inference network, parameterized by $\theta_D$, that represents a mapping $D: \mathbb{R}^{N} \rightarrow \mathbb{R}$
of the input image ($\textbf{f}$ or $\hat{\textbf{f}}$) to a real-valued scalar. In the adversarial process, $D$ is trained to maximally differentiate the synthetic image $\textbf{f}$ from the true image $\hat{\textbf{f}}$, and $G$ is trained to maximally fool $D$ such that the generated synthetic image $\hat{\textbf{f}}$ is wrongly classified as a true image. This adversarial process can be represented by a two-player minimax game with value function $V(D, G)$:
\begin{equation} \label{eq:GAN}
\min_{\theta_G} \max_{\theta_D} V(D,G)={E_{\textbf{f} \sim p_\textbf{f}}} [l\left(D(\textbf{f}; \theta_D)\right)] + {E_{z\sim p_z}} [l(1- D\left(G(\textbf{z}; \theta_G)\right) )],
\end{equation}
where $l(.)$ represents a suitable objective function. When the global optimum of this minimax game is achieved, the synthetic images generated by the generator $G$ can not be differentiated from the true images by using any observer, and the synthetic image distribution $p_{\hat{\textbf{f}}}$ equals the true image distribution $p_\textbf{f}$:  $p_{\hat{\textbf{f}}} = p_\textbf{f}$. Let $\theta_G^*$ and $\theta_D^*$ denote the optimal parameters for $G$ and $D$ respectively after stable convergence has been reached in the above minimax game.\\

\subsubsection{Progressive Growing of GANs (ProGAN)} 
\begin{figure}[htp]
	\centering
	{\includegraphics[width=0.8\linewidth]{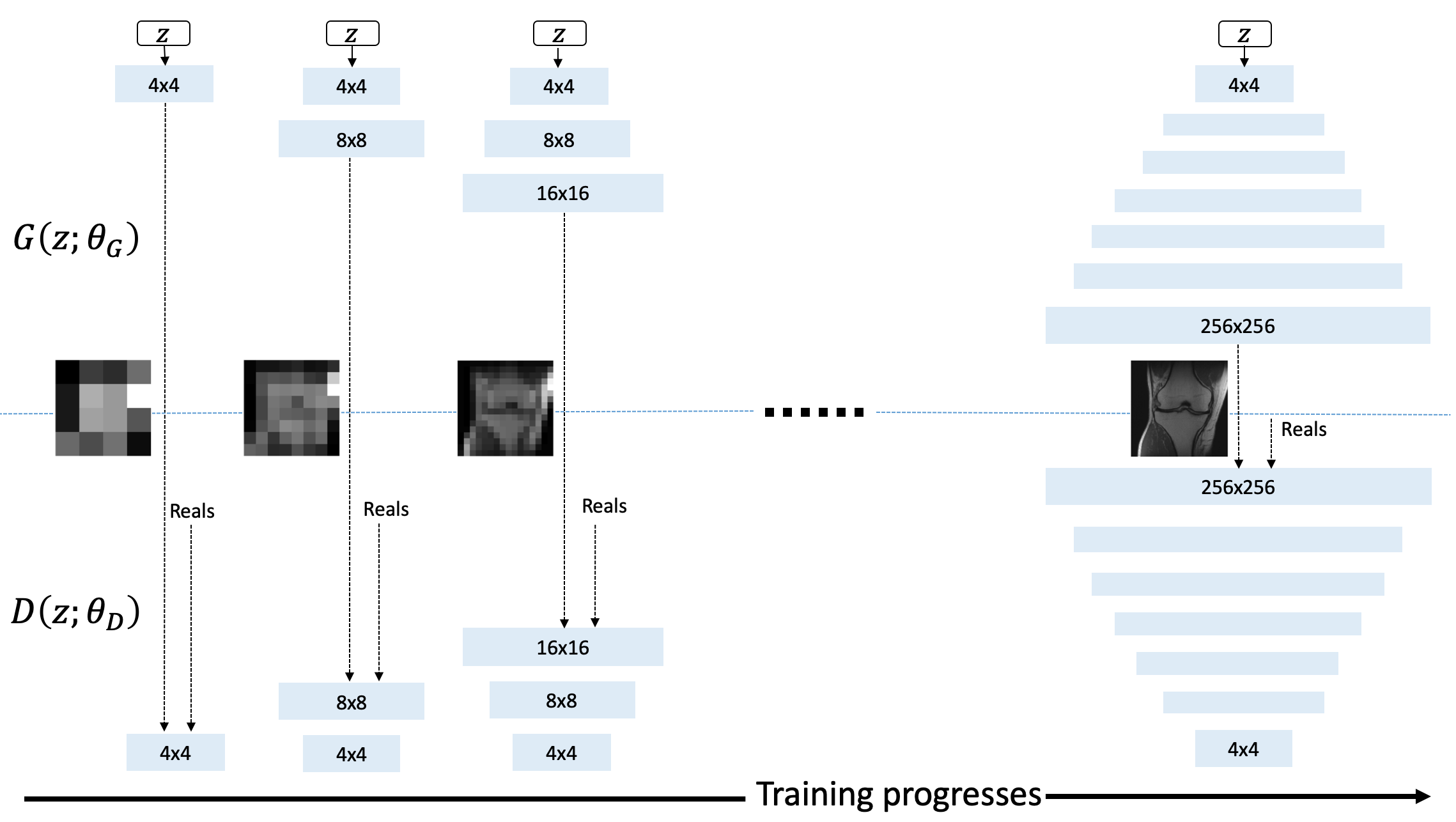}}
	\vspace{0.2cm}
	\caption{ProGAN: Training starts with generator $G$ and discriminator $D$ corresponding to low spatial resolution of $4\times4$ pixels. As training progresses, layers are added to $G$ and $D$ to gradually increase the spatial resolution of the generated images towards the final resolution, which for our study is $256\times256$.}
	\label{fig:PGGAN}
\end{figure}

In practice, however, stabilization of GAN training has been known to be difficult \cite{deeplearningbook2016}. This is primarily due to the unstable nature of the adversarial learning process, which involves the generator and the discriminator being trained simultaneously with competing objectives. This has served as a bottleneck in using GANs to reliably generate high-resolution images. Recently, Karras \textit{et al.} \cite{karras2018PGGAN} proposed a training strategy for GANs that has mitigated the stabilization problem of GAN training to great effect and resulted in GANs being able to generate realistic natural images at resolutions as high as $1024\times1024$ pixels. In this novel learning strategy called Progressive Growing of GANs (ProGAN), the training starts from low-resolution images and layers are added progressively to both the generator and the discriminator networks to increase the resolution (Fig. \ref{fig:PGGAN}). Such a progressive training strategy has resulted in higher stability in the training process for GANs and improved quality in the generated images, making ProGAN the current state-of-the-art method for training GANs on image data.

\subsection{Image-adaptive GAN-based reconstruction (IAGAN) for medical imaging}
Once a GAN has been stably trained and can generate images similar to samples from the true data distribution, the learned generator network can be used as a prior for solving linear inverse problems. In the context of medical imaging, the learned prior may be employed for reconstructing images from incomplete and/or corrupted measurement data. Given a pre-trained generative model, Bora \textit{et al.} \cite{bora2017csgm} proposed to find a vector in the latent space of the generator such that the corresponding image agrees with the observed measurement data. This leads to the following minimization problem:
\begin{equation}\label{eq:csgm}
\hat{\textbf{z}} = \argmin_\textbf{z} ||\textbf{g}-\textbf{H}G(\textbf{z};\theta_G^*)||^2_2,
\end{equation}
where the final reconstructed image is formed as $\hat{\textbf{f}}=G(\hat{\textbf{z}},\theta_G^*)$. However, it is difficult in practice to train perfect generators, and as a result, the generative model may not be able to span the entire range of the actual object distribution. This may lead to a lack of fidelity between the reconstruction obtained and the observed measurements. In order to mitigate this problem, Hussein \textit{et al.} \cite{hussein2019IAGAN} proposed an image-adaptive GAN-based (IAGAN) reconstruction framework where the image is constrained to lie in the range of $G$ while at the same time the trained generator's parameter weights are further tuned to enforce consistency with the observed measurement data. With the parameters of $G$ now denoted by $\theta$ and initialized as  $\theta^{(0)}=\theta_G^*$, $\textbf{z}$ and $\theta$ are jointly estimated as:
\begin{equation}\label{eq:IAGAN}
\hat{\textbf{z}}, \hat{\theta} = \argmin_{\textbf{z},\theta} ||\textbf{g}-\textbf{H}G(\textbf{z};\theta)||^2_2,
\end{equation}
where the final reconstructed image is formed as $\hat{\textbf{f}}=G(\hat{\textbf{z}};\hat{\theta})$. The authors also proposed to initialize the latent vector $\textbf{z}$ with the optimal solution obtained from solving Eq.\eqref{eq:csgm} to better condition the optimization problem. Since the generator network is differentiable, any suitable stochastic gradient-based method may be applied to solve Eq.\eqref{eq:IAGAN}. 

Additionally, regularization on the generative model in the form of a sparsity-promoting penalty may be added to the IAGAN framework to further mitigate artifacts resulting from data incompleteness and/or when the measurements contain a high level of noise. The optimization problem in Eq.\eqref{eq:IAGAN} may be modified as follows:

\begin{equation}\label{eq:IAGAN-TV}
\hat{\textbf{z}}, \hat{{\theta}} = \argmin_{\textbf{z},\theta} ||\textbf{g}-\textbf{H}G(\textbf{z};\theta)||^2_2 + \lambda \Phi(G(\textbf{z}; \theta)),
\end{equation}
with $\theta^{(0)}=\theta_G^*$, where $\Phi(.)$ is a suitable sparsity-promoting penalty function and $\lambda$ is a hyperparameter that controls the strength of regularization. In our experiments, we consider $\Phi(.)$ to be the total variation (TV) semi-norm \cite{barrett2013foundations}, and we refer to the method represented by Eq.\eqref{eq:IAGAN-TV} as IAGAN-TV. In this way, traditional sparsity-promoting regularization may be combined with GAN-constrained solutions, which may potentially enhance the quality of the reconstructed image as compared with using either method alone.

\section{Numerical Studies}
\label{sec:sections}
\subsection{Imaging system} 
In this preliminary study, the MR imaging system contains a single-coil with the forward operator $\textbf{H}$ corresponding to the discrete Fourier transform (DFT). Emulated single-coil (ESC) \cite{tygertESC2019} data from multi-coil acquisitions were employed for the reconstruction experiments as a first step to demonstrate the proposed method without involving the complexity of multiple receiver coils.

\subsection{Dataset} Zbontar \textit{et al.} \cite{fastMRI2018} recently released an open dataset called NYU fastMRI which contains a large number of raw MR k-space measurements as well as clinical MR images of human knees. The dataset includes both raw multi-coil and the corresponding ESC k-space data for 1594 knee volumes across different standard clinical MR systems and pulse sequences. The volumes are split into separate training, validation, test and challenge datasets. 
In our study, the ProGAN was trained using images obtained from the training dataset, and the IAGAN reconstruction method was applied on retrospectively subsampled full k-space data from the validation dataset. The test and challenge datasets were not considered for this study as they do not contain full k-space measurements, and hence ground truth images which can serve as gold standard reference can not be obtained for these datasets. 
The training images with full field-of-view (FOV) were obtained by performing the root sum-of-squares reconstruction method \cite{roemerRSS1990} over the centrally cropped $256\times256$ fully sampled region from the multi-coil k-space data in the training dataset. Excluding the first three noisy slices for each volume, this resulted in 31,823 training images with a spatial resolution of $256\times256$ pixels. Each training image slice was normalized by the maximum intensity in the corresponding volume.

\subsection{ProGAN training details}
For training the ProGAN, the code published by Karras \textit{et al.} at \url{https://github.com/tkarras/progressive_growing_of_gans} and implemented in Tensorflow \cite{abadi2016tensorflow} was employed. The default settings for the training parameters were used for our experiments. The training was performed on a system with an Intel Xeon E5-2620v4 Central Processing Unit (CPU) @ 2.1 GHz and 4 NVIDIA TITAN X Graphical Processing Units (GPUs).

\subsection{Baseline and reconstruction details}
\begin{figure}[htp]
	\centering
	\includegraphics[width=0.3\linewidth]{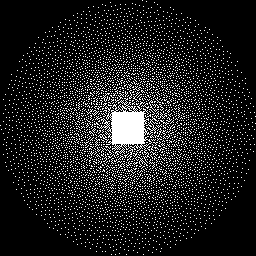}
	\vspace{0.3cm}
	\caption{Variable-density Poisson-disc sampling pattern in k-space with acceleration factor $R=8$. The central $32\times32$ region of the k-space is fully sampled.}
	\label{fig:mask}
\end{figure}

The IAGAN framework was employed to reconstruct images from undersampled ESC k-space data from a slice in a volume in the validation dataset. A volume obtained with the coronal proton density without fat suppression (CORPD) data-acquisition protocol was considered in our reconstruction experiments. The ground truth for comparing reconstructions was obtained by performing the inverse discrete Fourier transform (IDFT) of the fully sampled ESC k-space data.  Variable-density Poisson-disc sampling \cite{sparseMRI} with an acceleration factor of $R=8$ was used to undersample the full raw k-space data (Fig. \ref{fig:mask}). The sampling pattern was generated using the Berkeley Advanced Reconstruction Tolbox (BART) \cite{BART}. Zero-filled (ZF) reconstruction refers to the IDFT of the k-space zero-filled after subsampling and contains severe aliasing artifacts. As a reference, we consider a penalized least squares solution with TV penalty (PLS-TV) as in Eq.\eqref{eq:pls}, obtained by using the BART toolbox. The hyperparameter $\lambda$ was chosen by peforming a grid search and selecting the value that resulted in the lowest mean square error (MSE) of the reconstructed image with respect to the ground truth. Images were reconstructed with the CSGM, IAGAN and IAGAN-TV methods using Tensorflow \cite{abadi2016tensorflow}. The Adam optimizer \cite{kingma2014adam} was used to perform stochastic gradient descent for solving Eq.\eqref{eq:csgm}, Eq.\eqref{eq:IAGAN} and Eq.\eqref{eq:IAGAN-TV} respectively.

MR images have both magnitude and phase components. Since the reconstructions are not being performed directly from multi-coil acquisitions but rather from the corresponding emulated single-coil data, the phase information cannot be recovered with reasonable accuracy. Hence, for this preliminary study, the phase information from the fully sampled ESC k-space data was retrospectively added into all the considered reconstruction methods with subsampled ESC data.

\section{Results}
\subsection{ProGAN training results}
\begin{figure}[H]
	\centering
	\includegraphics[width=1.0\linewidth]{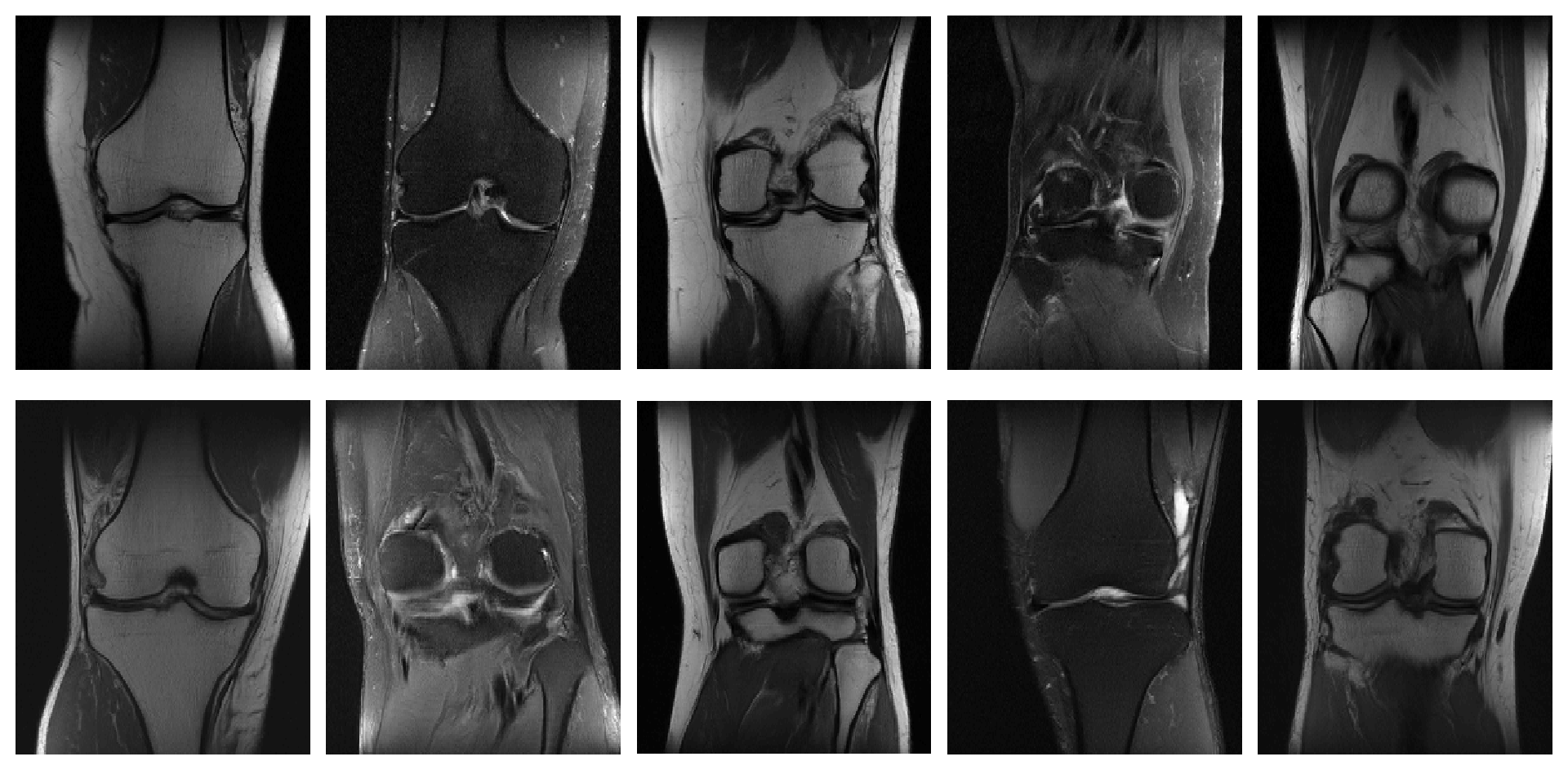}
	\caption{(Top) Samples from knee MRI images in the training dataset (Bottom) Samples from generated knee MRI images after training with the ProGAN}
	\label{fig:real_fake_samples}
\end{figure}

After the ProGAN was trained, the images produced by the generator highly resembled the true knee images in the training dataset. For visual comparison, samples of images from the training dataset as well as samples of images generated by the ProGAN after training, cropped to the central region of interest (ROI), are shown in Fig. \ref{fig:real_fake_samples}.

\subsection{Reconstruction from undersampled validation k-space data using IAGAN}
\begin{figure}[htp]
	\centering
	\includegraphics[width=1.0\linewidth]{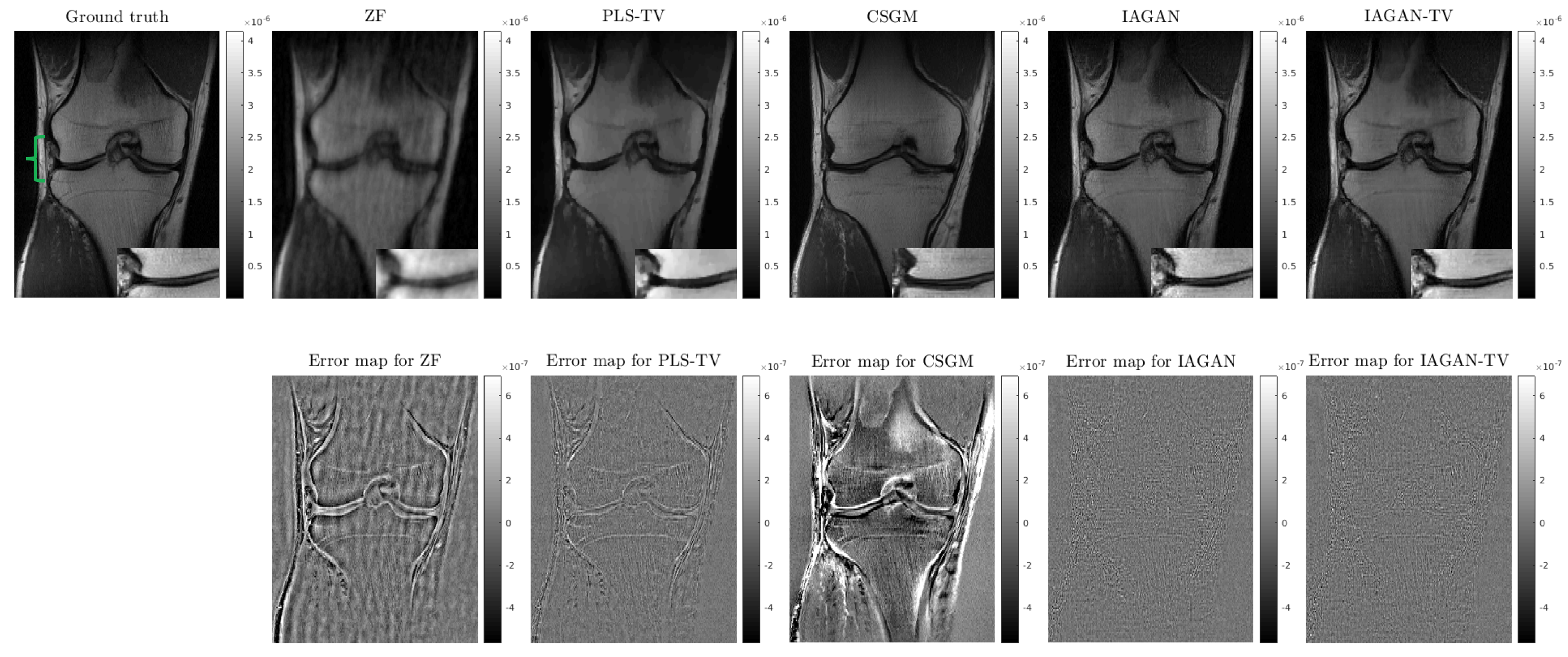}
	\vspace{0.1cm}
	\caption{(Top row) \textit{from left to right}: Ground truth, ZF reconstruction, PLS-TV reconstruction, CSGM reconstruction, IAGAN reconstruction and IAGAN-TV reconstruction. The meniscular region indicated by the green bracket in the ground truth image is expanded for each reconstruction result at the bottom right corner of each image. (Bottom row) \textit{from left to right}: Error maps for ZF, PLS-TV, CSGM, IAGAN and IAGAN-TV reconstructions respectively.}
	\label{fig:subplot_results}
\end{figure}

Images reconstructed by use of the IAGAN and the IAGAN-TV methods are compared with the baseline PLS-TV algorithm as well as the CSGM method in Fig. \ref{fig:subplot_results}.
 From the results, it can be observed that the IAGAN reconstruction contains fine details in regions that are of potential interest for medical diagnosis and retains the bone texture to a large degree, as compared with the PLS-TV method where these relevant features could not be reliably recovered. This can be further highlighted with an expanded view of the lateral meniscus for each reconstructed image. Radiologists rely on a clear view of the meniscular region in order to detect tears in the knee \cite{vohra2011knee}. However, it can be observed that the meniscus appears oversmoothed in the PLS-TV reconstruction, while for the IAGAN reconstruction, it remains sharp with discernible fine features. On the other hand, the image produced by the CSGM method contains detailed features and texture information similar to knee images in the training data, but lacks fidelity with the observed measurements. The IAGAN-TV reconstruction regularizes the IAGAN solution and can remove some of the grain-like artifacts that may appear due to data incompleteness. The error maps with respect to the ground truth further illustrate the points above.

\section{Conclusion}
This study demonstrates the use of an image-adaptive GAN-based (IAGAN) algorithm to reconstruct high fidelity MR images from noisy and/or incomplete measurement data. A ProGAN was trained on a publicly available knee MRI dataset and the learned generator was employed in the IAGAN framework to reconstruct images from emulated single-coil raw data. Reconstructed images illustrate that the IAGAN method can recover fine features with diagnostic relevance in the image which may be oversmoothed by traditional sparsity-based reconstruction methods. Moreover, the IAGAN algorithm maintains data consistency while the CSGM method fails to maintain fidelity with the observed measurements, which is critical for medical diagnosis. In the context of MRI, it will be important to extend the current implementation to reconstruct images from subsampled multi-coil raw data, as well as investigate the impact of IAGAN-based reconstruction across different patient volumes, slices and data-acquisition protocols. Comparison of the IAGAN method with other recent deep learning-based reconstruction solutions for accelerated MRI, \cite{yang2016admmnet,hammernik2018vnn} as well as existing GAN-based reconstruction approaches \cite{quan2017cyclicGANMRI,mardani2019gancs} will be studied in the future.  Further, the IAGAN may be implemented in other imaging modalities such as low-dose x-ray CT, where such a GAN-constrained reconstruction method may help to reduce the amount of ionizing radiation in the patient's body. Finally, there remains a need to perform reader studies as well as quantitative analysis of the IAGAN-based reconstruction method.

\section*{Disclosure}       
This work is original and has not been submitted for publication or presentation elsewhere.

\bibliography{ReconGAN.bib}

\begin{thebibliography}{10}

\bibitem{barrett2013foundations}
Barrett, H.~H. and Myers, K.~J.,  [{\em Foundations of {I}mage
  {S}cience}{\nolinebreak\hspace{0.1em}]}, John Wiley \& Sons (2013).

\bibitem{rudin1992TV}
Rudin, L.~I., Osher, S., and Fatemi, E., ``Nonlinear total variation based
  noise removal algorithms,'' {\em Physica D: Nonlinear Phenomena}~{\bf 60}(1),
   259 -- 268 (1992).

\bibitem{lustig2007CS}
Lustig, M., Donoho, D.~L., Santos, J.~M., and Pauly, J.~M., ``Compressed
  sensing {MRI},'' in [{\em IEEE SIGNAL PROCESSING
  MAGAZINE}{\nolinebreak\hspace{0.1em}]},  (2007).

\bibitem{tian2011low}
Tian, Z., Jia, X., Yuan, K., Pan, T., and Jiang, S.~B., ``Low-dose {CT}
  reconstruction via edge-preserving total variation regularization,'' {\em
  Physics in Medicine \& Biology}~{\bf 56}(18),  5949 (2011).

\bibitem{goodfellow2014generative}
Goodfellow, I., Pouget-Abadie, J., Mirza, M., Xu, B., Warde-Farley, D., Ozair,
  S., Courville, A., and Bengio, Y., ``Generative adversarial nets,'' in [{\em
  Advances in Neural Information Processing
  Systems}{\nolinebreak\hspace{0.1em}]},   2672--2680 (2014).

\bibitem{quan2017cyclicGANMRI}
Quan, T.~M., Nguyen{-}Duc, T., and Jeong, W., ``Compressed sensing {MRI}
  reconstruction with cyclic loss in generative adversarial networks,'' {\em
  CoRR}~{\bf abs/1709.00753} (2017).

\bibitem{mardani2019gancs}
{Mardani}, M., {Gong}, E., {Cheng}, J.~Y., {Vasanawala}, S.~S., {Zaharchuk},
  G., {Xing}, L., and {Pauly}, J.~M., ``Deep generative adversarial neural
  networks for compressive sensing {MRI},'' {\em IEEE Transactions on Medical
  Imaging}~{\bf 38},  167--179 (Jan 2019).

\bibitem{bora2017csgm}
Bora, A., Jalal, A., Price, E., and Dimakis, A.~G., ``Compressed sensing using
  generative models,'' in [{\em Proceedings of the 34th International
  Conference on Machine Learning - Volume 70}{\nolinebreak\hspace{0.1em}]},
  {\em ICML'17},  537--546, JMLR.org (2017).

\bibitem{hussein2019IAGAN}
{Abu Hussein}, S., {Tirer}, T., and {Giryes}, R., ``{Image-Adaptive GAN based
  Reconstruction},'' {\em arXiv e-prints} ,  arXiv:1906.05284 (Jun 2019).

\bibitem{karras2018PGGAN}
Karras, T., Aila, T., Laine, S., and Lehtinen, J., ``Progressive growing of
  {GAN}s for improved quality, stability, and variation,'' {\em CoRR}~{\bf
  abs/1710.10196} (2017).

\bibitem{deeplearningbook2016}
Goodfellow, I., Bengio, Y., and Courville, A.,  [{\em Deep
  Learning}{\nolinebreak\hspace{0.1em}]}, The MIT Press (2016).

\bibitem{tygertESC2019}
{Tygert}, M. and {Zbontar}, J., ``{Simulating single-coil MRI from the
  responses of multiple coils},'' {\em arXiv e-prints} ,  arXiv:1811.08026 (Nov
  2018).

\bibitem{fastMRI2018}
Zbontar, J., Knoll, F., Sriram, A., Muckley, M.~J., Bruno, M., Defazio, A.,
  Parente, M., Geras, K.~J., Katsnelson, J., Chandarana, H., Zhang, Z.,
  Drozdzal, M., Romero, A., Rabbat, M., Vincent, P., Pinkerton, J., Wang, D.,
  Yakubova, N., Owens, E., Zitnick, C.~L., Recht, M.~P., Sodickson, D.~K., and
  Lui, Y.~W., ``fast{MRI}: An open dataset and benchmarks for accelerated
  {MRI},'' {\em CoRR}~{\bf abs/1811.08839} (2018).

\bibitem{roemerRSS1990}
Roemer, P.~B., Edelstein, W.~A., Hayes, C.~E., Souza, S.~P., and Mueller,
  O.~M., ``The {NMR} phased array,'' {\em Magnetic Resonance in Medicine}~{\bf
  16}(2),  192--225 (1990).

\bibitem{abadi2016tensorflow}
Abadi, M., Barham, P., Chen, J., Chen, Z., Davis, A., Dean, J., Devin, M.,
  Ghemawat, S., Irving, G., Isard, M., et~al., ``Tensorflow: a system for
  large-scale machine learning.,'' in [{\em OSDI}{\nolinebreak\hspace{0.1em}]},
    {\bf 16},  265--283 (2016).

\bibitem{sparseMRI}
Lustig, M., Donoho, D., and Pauly, J.~M., ``Sparse {MRI}: The application of
  compressed sensing for rapid {MR} imaging,'' {\em Magnetic Resonance in
  Medicine}~{\bf 58}(6),  1182--1195 (2007).

\bibitem{BART}
Uecker, M., Tamir, J., Ong, F., Holme, C., and Lustig, M., ``{BART}: version
  0.4.01,'' (June 2017).

\bibitem{kingma2014adam}
Kingma, D.~P. and Ba, J., ``Adam: A method for stochastic optimization,'' {\em
  arXiv preprint arXiv:1412.6980}  (2014).

\bibitem{vohra2011knee}
Vohra, S., Arnold, G., Doshi, S., and Marcantonio, D., ``Normal {MR} imaging
  anatomy of the knee,'' {\em Magnetic {R}esonance {I}maging clinics of North
  America}~{\bf 19},  637--53, ix (08 2011).

\bibitem{yang2016admmnet}
Yang, Y., Sun, J., Li, H., and Xu, Z., ``Deep {ADMM-N}et for compressive
  sensing {MRI},'' in [{\em Advances in Neural Information Processing Systems
  29}{\nolinebreak\hspace{0.1em}]},  Lee, D.~D., Sugiyama, M., Luxburg, U.~V.,
  Guyon, I., and Garnett, R., eds.,  10--18, Curran Associates, Inc. (2016).

\bibitem{hammernik2018vnn}
Hammernik, K., Klatzer, T., Kobler, E., Recht, M.~P., Sodickson, D.~K., Pock,
  T., and Knoll, F., ``Learning a variational network for reconstruction of
  accelerated {MRI} data,'' {\em Magnetic Resonance in Medicine}~{\bf 79}(6),
  3055--3071 (2018).

\end{thebibliography}
\bibliographystyle{spiebib} 

\end{document}